\documentclass{osa-article}

\journal{osajournal}
\usepackage{multirow}
\usepackage{bm}

\articletype{article}
\begin{document}
\title{Demonstration of 4H silicon carbide on aluminum nitride integrated photonics platform}

\author{Ruixuan Wang\authormark{1}, Jingwei Li\authormark{1}, Lutong Cai \authormark{1}, and Qing Li\authormark{*1}}
\address{\authormark{1}Department of Electrical and Computer Engineering, Carnegie Mellon University, Pittsburgh, PA 15213, USA}
\email{\authormark{*}qingli2@andrew.cmu.edu}
\begin{abstract}
The existing silicon-carbide-on-insulator photonic platform utilizes a thin layer of silicon dioxide under silicon carbide to provide optical confinement and mode isolation. Here, we replace the underneath silicon dioxide layer with a 1-$\mu$m-thick aluminum nitride and demonstrate a 4H-silicon-carbide-on-aluminum-nitride integrated photonics platform for the first time. Efficient grating couplers, low-loss waveguides, and compact microring resonators with intrinsic quality factors up to 210,000 are fabricated. In addition, by undercutting the aluminum nitride layer, the intrinsic quality factor of the silicon carbide microring is improved by nearly one order of magnitude (1.8 million). Finally, an optical pump-probe method is developed to measure the thermal conductivity of the aluminum nitride layer, which is estimated to be over 30 times of that of silicon dioxide.  
\end{abstract}

\noindent Integrated photonic platforms such as silicon-on-insulator (SiOI) and lithium-niobate-on-insulator (LNOI) have revolutionized modern optical technologies \cite{SOI_review, LNOI_reivew, Review_nphoton_Integrated}. In addition to these mature device platforms, there are emergent candidates such as silicon carbide (SiC) which find potential applications in both classical and quantum domains \cite{Vuckovic_SiC_review}. Traditionally, the insulator material consists of a thin layer of silicon dioxide, which serves as a lower index cladding (index $\approx 1.45$ in the 1550 nm) to the core so that the optical mode is sufficiently confined and isolated from the substrate. However, silicon dioxide suffers from strong optical absorption in the mid-infrared, which increases the propagation loss of the optical mode at longer wavelengths \cite{Lipson_SOI_midIR}. In addition, it has a much smaller thermal conductivity compared to other photonic materials such as sapphire or aluminum nitride (AlN). As a result, heat transfer from the photonic core to the substrate is typically slow and inefficient, often leading to significant thermo-optic bistability when the photonic device is operated under medium to high optical powers \cite{Vahala_thermal}. 

To extend the material transparency to the mid-infrared as well as to mitigate the thermal effect of integrated photonic devices, wide-bandgap materials such as sapphire have been suggested to replace the silicon dioxide layer. Examples include Si-on-sapphire \cite{SionSa}, AlN-on-sapphire\cite{Wang_AlNonSa}, and LN-on-sapphire\cite{Safavi-LNonSa}. In this work, we propose an alternative combination, that is, to employ AlN as the underlying cladding to SiC, and demonstrate a 4H-SiC-on-AlN integrated photonic platform for the first time. The choice of AlN for the 4H-SiCOI platform is motivated by several factors: first, AlN is transparent from ultraviolet to the mid-infrared band \cite{Tang_AlN_UVQ, Guo_comb_AlN}; second, AlN, in particular single-crystal AlN, has a much larger thermal conductivity compared to silicon dioxide \cite{AlN_thermal_conductivity}; and finally, the lattice constant of AlN is closely matched to that of 4H-SiC, offering material compatibility between these two layers \cite{Growth_AlNonSiC}. 
\begin{figure}[ht]
\centering
\includegraphics[width=0.85\linewidth]{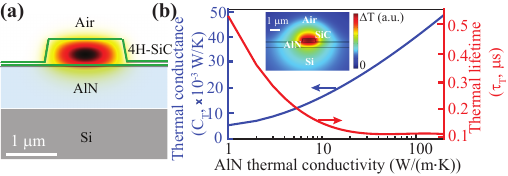}
\caption{(a) Schematic of the 4H-silicon-carbide-on-aluminum-nitride integrated photonics platform. (b) Simulated thermal conductance (blue solid line) and lifetime (red solid line) for a 50-$\mu$m-radius 4H-SiC microring as a function of thermal conductivity of the underlying 1-$\mu$m-thick aluminum nitride layer. The SiC ring width is $2.5\ \mu$m, and the SiC layer has a height of 700 nm with an etch depth of $550$ nm. The inset shows the temperature distribution corresponding to a thermal conductivity of 100 $\text{W}/(\text{m}\cdot\text{K})$.}
\label{Fig1}
\end{figure}

The mode profile of a representative 4H-SiC-on-AlN waveguide is provided in Fig.~1(a). The refractive indices of 4H-SiC and AlN in the 1550 nm band are around $2.6$ and $2.1$, respectively. With this modest index contrast, the substrate leakage loss of the fundamental transverse-electric (TE) mode is estimated to be near 1 dB/cm (1 dB/m) with a $1\ \mu$m ($1.5\ \mu$m) thick AlN layer. To quantify the improvement of the thermal properties brought by AlN, we use a 50-$\mu$m-radius SiC microring as an example in Fig.~1(b). As can be seen, when the thermal conductivity of the under clad is increased from $1\ \text{W}/(\text{m}\cdot\text{K})$ (typical of silicon dioxide) to $>100\ \text{W}/(\text{m}\cdot\text{K})$ (typical of AlN), the thermal conductance ($C_T$) of the corresponding microring is enhanced by a factor of more than 7 while the thermal lifetime ($\tau_T$) is reduced by a factor of more than 5. Here, the thermal conductance of a microresonator is defined as $C_T\equiv P_{abs}/\Delta T$, where $P_{abs}$ and $\Delta T$ represent the absorbed power and the increased temperature in the optical mode area, respectively. The thermal lifetime, on the other hand, denotes the time constant when an exponential function is used to fit the dynamical change of the temperature (i.e.,$\exp(-t/\tau_T)$). Hence, the simulation data in Fig.~1(b) suggests that the thermal effect in the SiCOI platform can be strongly suppressed by replacing the underlying silicon dioxide with AlN. This is particularly meaningful to nonlinear optical applications, where the optical pump is partially absorbed by the microresonator and the resulting thermo-optic bistability hinders applications such as soliton microcomb generation \cite{Li_SiN_octave}. In contrast, a substantial increase of the thermal conductance of the microresonator can lead to direct landing on the soliton state without auxiliary thermal mitigation \cite{Li_SiN_octave, Wong_Aux_laser_soliton}.

\begin{figure}[ht]
\centering
\includegraphics[width=0.8\linewidth]{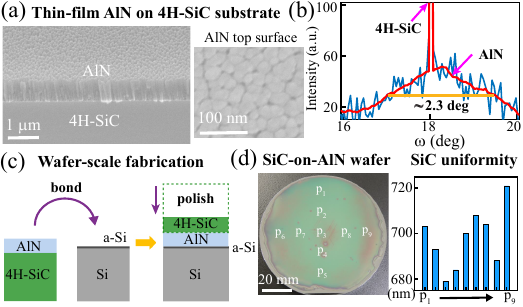}
\caption{(a) Cross-sectional (left) and top surface (right) view of scanning electron micrograph of 1-$\mu$m-thick AlN sputtered on a 500-$\mu$m-thick 4H-SiC wafer. (b) Rocking curve measurement of the AlN layer reveals a full-width at half maximum angle around $2.3$ degrees (the adjacent narrow peak is from 4H-SiC). (c) Illustration of a customized bonding and polishing process to fabricate SiC-on-AlN wafers. (d) Optical micrograph of a 4-inch-size SiC-on-AlN wafer (left) and the measured SiC thickness across the wafer area (right).}
\label{Fig2}
\end{figure}

To fabricate SiC-on-AlN wafers, we first sputter a 1-$\mu$m-thick AlN on a semi-insulating, 500-$\mu$m-thick 4H-SiC wafer (Kyma technologies). Increasing the AlN thickness to more than 1 $\mu$m on 4H-SiC turns out to be technically challenging, as the internal stress built inside the AlN layer becomes strong enough to cause cracks. The scanning electron micrographs (SEMs) for the cross section and top surface of the AlN layer confirm a characteristic columnar structure, with the AlN grain size estimated to be between 15 nm to 30 nm (Fig.~2(a)). The material quality of the AlN layer is further examined by performing a rocking curve measurement using x-ray diffraction, revealing a full-width at half maximum angle of $2.3$ degrees (Fig.~2(b)). Next, we proceed to bond the AlN-on-SiC wafer to a Si substrate using amorphous Si as the bonding layer, and the top SiC layer is subsequently polished down to less than $1\ \mu$m (Fig.~2(c), NGK Insulators). Note that this bonding and polishing process is similar to what has been employed for the fabrication of conventional SiC-on-oxide wafers \cite{Lin_3CSiC, Adibi_3CSiC, Noda_4HSiC_PhC, Vuckovic_4HSiC_nphoton, Ou_4HSiC_combQ, Li_4HSiC_comb}. Finally, Fig.~2(d) shows the optical micrograph of a 4-in-size SiC-on-AlN wafer, for which the variation of the SiC thickness is found to be less than $3\%$ across the entire wafer (with a nominal SiC thickness around 700 nm).

The SiC-on-AlN wafer is further characterized by fabricating chip-scale optical devices using an optimized nanofabrication process for 4H-SiC \cite{Li_4HSiC_comb}. Figure 3(a) shows one such example, which is a grating-coupled 50-$\mu$m-radius SiC microring (ring width of $2.5\ \mu$m). With a nominal SiC thickness of 700 nm and an etch depth of 550 nm, the grating coupler exhibits a peak coupling efficiency of $15\%$-$20\%$ (7-8 dB). Its 3-dB coupling bandwidth is estimated from the laser transmission to be more than 60 nm (Fig.~3(b)). In addition, we extract the intrinsic $Q$ of the fundamental TE mode to be near $0.21$ million (Fig.~3(c)), which corresponds to a propagation loss of $1.7$ dB/cm. Since this level of $Q$ is considerably lower than the $Q$s obtained from 4H-SiC-on-oxide devices (where intrinsic $Q$s are typically larger than 1 million) \cite{Li_4HSiC_comb}, it is likely limited by the leakage loss to the substrate due to insufficient isolation (simulated leakage loss is around 1 dB/cm for 1 $\mu$m AlN). In addition, there could be additional scattering and absorption losses introduced by the AlN layer, which has a relatively large grain size. Nevertheless, achieving a propagation loss on the order of $1$ dB/cm for the SiC-on-AlN wafer demonstrates its viability for low-loss photonic applications.

\begin{figure}[ht]
\centering
\includegraphics[width=0.8\linewidth]{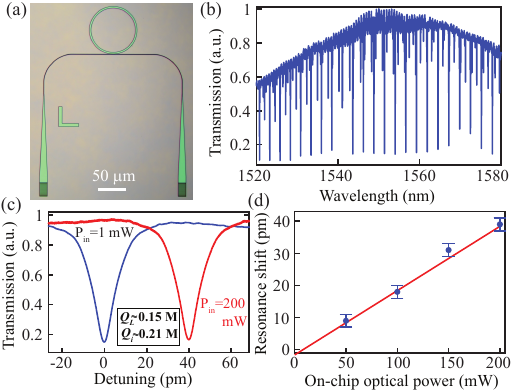}
\caption{(a) Optical micrograph of a grating coupled 4H-SiC microring resonator with $50\ \mu$m radius. (b) Optical transmission of the microring. (c) Representative resonance of the fundamental transverse-electric mode under two different optical powers, displaying an intrinsic $Q$ around 210k and an approximate thermo-optic shift of 40 pm for 200 mW on-chip power. (d) Experimentally measured thermal shifts of the resonance (markers with error bars) and their linear fitting (red solid line) as a function of the on-chip optical power.}
\label{Fig3}
\end{figure}

At increased pump powers, the SiC resonance only experiences a thermally induced resonance shift without exhibiting any thermo-optic bistability (an approximate 40 pm shift is observed for 200 mW input power, see Fig.~3(c)). Such wavelength shifts are found to to be linearly proportional to the optical power (Fig.~3(d)). Using the thermo-optic coefficient of 4H-SiC ($dn/dT\approx 4.2\times 10^{-5}$/K) \cite{Ou_4HSiC_thermal}, the resonance is expected to shift 25 pm/K with the temperature change. Combining these numbers, we formulate that
\begin{equation}
\Delta T=\frac{\eta P_{in}}{C_T}= \frac{\eta \times 0.2\ \text{W}}{C_T}\approx \frac{40 \ \text{pm}}{25\ \text{pm}/\text{K}},
\end{equation}
where $\eta$ represents the percentage of the pump power absorbed by the microresonator. As a result, the thermal conductance of the 50-$\mu$m-radius SiC microring is estimated to be $C_T\approx (125\eta \times 10^{-3})$ W/K.  While it is difficult to know the precise value of $\eta$ for our device, using a conservative estimation of $20\%$ for $\eta$ points to a thermal conductance of $(25\times 10^{-3})$ W/K. By comparing this result to the simulation data in Fig.~1(b), we infer that the thermal conductivity of the AlN layer is on the order of (20-30) $\text{W}/(\text{m}\cdot\text{K})$ .

\begin{figure}[ht]
\centering
\includegraphics[width=0.8\linewidth]{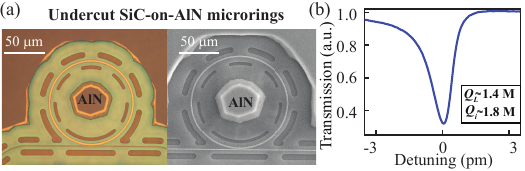}
\caption{(a) Optical micrograph (left) and scanning electron micrograph (right) of an undercut SiC microring. The SiC microring has a radius of $50\ \mu$m and the underneath AlN pedestal has an approximate radius of $20\ \mu$m. (b) Representative linear transmission of the undercut SiC microring, displaying an estimated intrinsic $Q$ around $1.8$ million.}
\label{Fig4}
\end{figure}

To verify that the intrinsic $Q$ of SiC-on-AlN microrings is indeed limited by the substrate leakage loss and AlN-induced absorption/scattering losses, we open trenches surrounding the SiC microring and etch away AlN using AZ 400K solutions at $50^{\circ}C$. As shown in Fig.~4(a), the undercut SiC microring resonator is still mechanically robust owing to a 150 nm unetched SiC layer supporting the overall structure (except for the trench regions). With the removal of AlN, the intrinsic $Q$ of the SiC microring dramatically improves to $1.8$ million, which is almost one order of magnitude larger than the $0.21$ million $Q$ measured in Fig.~3(c). This difference suggests that the propagation loss for SiC-on-AlN devices can be well below 1 dB/cm if we manage to increase the AlN thickness and further improve its material quality \cite{Growth_AlNonSiC}.  

In the undercut structure, the heat generated inside the SiC microring travels to the AlN pedestal through the unetched SiC layer before dissipating to the Si substrate, which helps reduce the thermal lifetime. Hence, an accurate measurement of the thermal decay time enables an estimation of the thermal conductivity of the AlN layer without the knowledge of the absorption rate (i.e., $\eta$). For this purpose, we develop an optical pump-probe experimental scheme as illustrated in Fig.~5(a): First, a pump laser with a relatively strong optical power is employed, whose transmission scan displays a typical thermo-optic bistability when coupled to the microring resonance; second, a probe laser is coupled to a different resonance by a varied red detuning ($\Delta\lambda_\text{probe}$), and its optical transmission is monitored by a fast oscilloscope; third, a small wavelength tuning of the pump laser across the bistability point triggers the temperature cooling of the microresonator (trigger time marked as $t_0$), resulting a blue-shift of all the cavity resonances simultaneously; and finally, by monitoring the time that it takes for the blue-shifting resonance to coincide with the probe laser ($t_p$), we can infer the thermal lifetime of the microresonator. 

\begin{figure}[ht]
\centering
\includegraphics[width=0.8\linewidth]{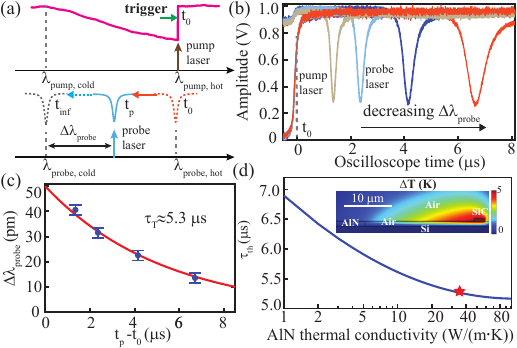}
\caption{(a) Illustration of a pump-probe scheme to measure the thermal lifetime of a microresonator: the probe laser has a preset cold-cavity detuning ($\Delta\lambda_p$) while the pump laser is slightly varied near the optical bistability point to trigger the thermo-optic wavelength scanning. (b) Superimposed traces of the probe laser transmission corresponding to varied wavelength detunings. (c) Experimentally measured waiting time $t_p-t_0$ for different $\Delta \lambda_\text{probe}$ (markers). The red solid line is the exponential fitting with a time constant of $5.3\ \mu$s. (d) Simulated thermal lifetime for the undercut 50-$\mu$m-radius SiC microring, which is compared to the experimental data (red star) for the estimation of AlN's thermal conductivity. The inset shows one simulation example for the temperature distribution in such undercut devices (the AlN pedestal is assumed to have a radius of 20 $\mu$m.)}
\label{Figure5}
\end{figure}

Figure 5(b) shows one example of the pump-probe measurement, in which a series of probe detunings ($\Delta\lambda_\text{probe}$) are tested for a fixed pump power and the corresponding probe traces are recorded. Note that while each trace appears to be a regular resonance scan, there is a major difference. In the regular resonance scan, the cavity resonance is fixed in wavelength and it is the laser detuning that is varied. In our case, however, the probe laser has a fixed detuning and it is the cooling of the cavity temperature (triggered by the pump laser's thermo-optic bistability) that leads to the self-scanning of the cavity resonance. This understanding also explains that while we are monitoring the same probe resonance, the apparent resonance width becomes larger as $\Delta\lambda_\text{probe}$ is reduced (Fig.~5(b)). This is because the thermo-optic scanning rate slows down as the temperature drops (exponential decay). By fitting the measured $t_p-t_0$ for various probe detuning $\Delta\lambda_\text{probe}$, we extract the thermal lifetime of the undercut SiC microring to be approximately $5.3\ \mu$s (Fig.~5(c)). In Fig.~5(d), by comparing this data to the simulated thermal lifetimes, the thermal conductivity of the AlN layer is estimated to be near 35 $\text{W}/(\text{m}\cdot\text{K})$, or approximately 30 times of that of silicon dioxide. Note that while this level of thermal conductivity is typical for amorphous AlN, which is the case here since our AlN is obtained from a sputtering process, a thermal conductivity of more than 300 $\text{W}/(\text{m}\cdot\text{K})$ has been reported in the literature for single-crystal AlN \cite{AlN_thermal_conductivity}. 

In conclusion, we propose and demonstrate a 4H-SiC-on-AlN integrated photonics platform for the first time. With 1-$\mu$m-thick AlN under a 700 nm SiC device layer, we fabricate microrings with intrinsic quality factors up to 210,000 in the 1550 nm band, which correspond to an on-chip propagation loss of $1.7$ dB/cm. By undercutting the AlN layer, the intrinsic quality factors have improved by nearly one order of magnitude ($1.8$ million). In addition, a pump-probe scheme based on the optically-induced bistability enables a direct estimation of the thermal conductivity of the sputtered AlN layer, which is approximately 30 times of that of silicon dioxide. With further improvement in the AlN growth technologies, we believe such SiC-on-AlN integrated photonics platform can support propagation losses well below 1 dB/cm from ultraviolet to mid-infrared, while offering excellent suppression of the thermal effect for a wide range of high-power applications.     

\begin{backmatter}
\bmsection{Funding}
This work was supported by DARPA (D19AP00033) and NSF (2127499). 

\bmsection{Acknowledgments}
The authors acknowledge the use of Bertucci Nanotechnology Laboratory at Carnegie Mellon University supported by grant BNL-78657879 and the Materials Characterization Facility supported by grant MCF-677785. R.~Wang also acknowledges the support of Tan Endowed Graduate Fellowship from CMU. 

\bmsection{Disclosures}  The authors declare no conflicts of interest.

\bmsection{Data Availability} Data underlying the results presented in this paper are not publicly available at this time but may be obtained from the authors upon reasonable request.

\end{backmatter}
\bibliography{SiC_Ref}

\begin{thebibliography}{10}
\newcommand{\enquote}[1]{``#1''}

\bibitem{SOI_review}
S.~Shekhar, W.~Bogaerts, L.~Chrostowski, J.~E. Bowers, M.~Hochberg, R.~Soref,
  and B.~J. Shastri, \enquote{Roadmapping the next generation of silicon
  photonics,} {\protect\JournalTitle{Nature Communications}} \textbf{15}, 751
  (2024). Number: 1 Publisher: Nature Publishing Group.

\bibitem{LNOI_reivew}
D.~Zhu, L.~Shao, M.~Yu, R.~Cheng, B.~Desiatov, C.~J. Xin, Y.~Hu, J.~Holzgrafe,
  S.~Ghosh, A.~Shams-Ansari, E.~Puma, N.~Sinclair, C.~Reimer, M.~Zhang, and
  M.~Lončar, \enquote{Integrated photonics on thin-film lithium niobate,}
  {\protect\JournalTitle{Advances in Optics and Photonics}} \textbf{13},
  242--352 (2021). Publisher: Optica Publishing Group.

\bibitem{Review_nphoton_Integrated}
J.~Wang, F.~Sciarrino, A.~Laing, and M.~G. Thompson, \enquote{Integrated
  photonic quantum technologies,} {\protect\JournalTitle{Nature Photonics}}
  \textbf{14}, 273--284 (2020). Number: 5 Publisher: Nature Publishing Group.

\bibitem{Vuckovic_SiC_review}
D.~M. Lukin, M.~A. Guidry, and J.~Vu{\v c}kovi{\'c}, \enquote{Integrated
  {{quantum photonics}} with {{silicon carbide}}: challenges and
  {{prospects}},} {\protect\JournalTitle{PRX Quantum}} \textbf{1}, 020102
  (2020).

\bibitem{Lipson_SOI_midIR}
S.~A. Miller, M.~Yu, X.~Ji, A.~G. Griffith, J.~Cardenas, A.~L. Gaeta, and
  M.~Lipson, \enquote{Low-loss silicon platform for broadband mid-infrared
  photonics,} {\protect\JournalTitle{Optica}} \textbf{4}, 707--712 (2017).
  Publisher: Optica Publishing Group.

\bibitem{Vahala_thermal}
T.~Carmon, L.~Yang, and K.~J. Vahala, \enquote{Dynamical thermal behavior and
  thermal self-stability of microcavities,} {\protect\JournalTitle{Optics
  Express}} \textbf{12}, 4742--4750 (2004). Publisher: Optica Publishing Group.

\bibitem{SionSa}
T.~Baehr-Jones, A.~Spott, R.~Ilic, A.~Spott, B.~Penkov, W.~Asher, and
  M.~Hochberg, \enquote{Silicon-on-sapphire integrated waveguides for the
  mid-infrared,} {\protect\JournalTitle{Optics Express}} \textbf{18},
  12127--12135 (2010). Publisher: Optica Publishing Group.

\bibitem{Wang_AlNonSa}
X.~Liu, C.~Sun, B.~Xiong, L.~Wang, J.~Wang, Y.~Han, Z.~Hao, H.~Li, Y.~Luo,
  J.~Yan, T.~Wei, Y.~Zhang, and J.~Wang, \enquote{Aluminum nitride-on-sapphire
  platform for integrated high-{Q} microresonators,}
  {\protect\JournalTitle{Optics Express}} \textbf{25}, 587--594 (2017).
  Publisher: Optica Publishing Group.

\bibitem{Safavi-LNonSa}
J.~Mishra, T.~P. McKenna, E.~Ng, H.~S. Stokowski, M.~Jankowski, C.~Langrock,
  D.~Heydari, H.~Mabuchi, M.~M. Fejer, and A.~H. Safavi-Naeini,
  \enquote{Mid-infrared nonlinear optics in thin-film lithium niobate on
  sapphire,} {\protect\JournalTitle{Optica}} \textbf{8}, 921--924 (2021).
  Publisher: Optica Publishing Group.

\bibitem{Tang_AlN_UVQ}
X.~Liu, A.~W. Bruch, Z.~Gong, J.~Lu, J.~B. Surya, L.~Zhang, J.~Wang, J.~Yan,
  and H.~X. Tang, \enquote{Ultra-high-{Q} {UV} microring resonators based on a
  single-crystalline {AlN} platform,} {\protect\JournalTitle{Optica}}
  \textbf{5}, 1279--1282 (2018).

\bibitem{Guo_comb_AlN}
H.~Weng, J.~Liu, A.~A. Afridi, J.~Li, J.~Dai, X.~Ma, Y.~Zhang, Q.~Lu, J.~F.
  Donegan, J.~F. Donegan, W.~Guo, and W.~Guo, \enquote{Directly accessing
  octave-spanning dissipative {{Kerr}} soliton frequency combs in an {{AlN}}
  microresonator,} {\protect\JournalTitle{Photonics Research}} \textbf{9},
  1351--1357 (2021).

\bibitem{AlN_thermal_conductivity}
A.~V. Inyushkin, A.~N. Taldenkov, D.~A. Chernodubov, E.~N. Mokhov, S.~S.
  Nagalyuk, V.~G. Ralchenko, and A.~A. Khomich, \enquote{On the thermal
  conductivity of single crystal {AlN},} {\protect\JournalTitle{Journal of
  Applied Physics}} \textbf{127}, 205109 (2020).

\bibitem{Growth_AlNonSiC}
C.-M. Zetterling, M.~Östling, K.~Wongchotigul, M.~G. Spencer, X.~Tang, C.~I.
  Harris, N.~Nordell, and S.~S. Wong, \enquote{Investigation of aluminum
  nitride grown by metal–organic chemical-vapor deposition on silicon
  carbide,} {\protect\JournalTitle{Journal of Applied Physics}} \textbf{82},
  2990--2995 (1997).

\bibitem{Li_SiN_octave}
Q.~Li, T.~C. Briles, D.~A. Westly, T.~E. Drake, J.~R. Stone, B.~R. Ilic, S.~A.
  Diddams, S.~B. Papp, and K.~Srinivasan, \enquote{Stably accessing
  octave-spanning microresonator frequency combs in the soliton regime,}
  {\protect\JournalTitle{Optica}} \textbf{4}, 193--203 (2017).

\bibitem{Wong_Aux_laser_soliton}
H.~Zhou, Y.~Geng, W.~Cui, S.-W. Huang, Q.~Zhou, K.~Qiu, and C.~Wei~Wong,
  \enquote{Soliton bursts and deterministic dissipative {Kerr} soliton
  generation in auxiliary-assisted microcavities,}
  {\protect\JournalTitle{Light: Science \& Applications}} \textbf{8}, 50
  (2019).

\bibitem{Lin_3CSiC}
X.~Lu, J.~Y. Lee, P.~X.-L. Feng, and Q.~Lin, \enquote{Silicon carbide microdisk
  resonator,} {\protect\JournalTitle{Optics Letters}} \textbf{38}, 1304--1306
  (2013).

\bibitem{Adibi_3CSiC}
T.~Fan, H.~Moradinejad, X.~Wu, A.~A. Eftekhar, and A.~Adibi,
  \enquote{High-{{Q}} integrated photonic microresonators on
  {{3C}}-{{SiC}}-on-insulator ({{SiCOI}}) platform,}
  {\protect\JournalTitle{Optics Express}} \textbf{26}, 25814--25826 (2018).

\bibitem{Noda_4HSiC_PhC}
B.-S. Song, T.~Asano, S.~Jeon, H.~Kim, C.~Chen, D.~D. Kang, and S.~Noda,
  \enquote{Ultrahigh-{{Q}} photonic crystal nanocavities based on {{4H}}
  silicon carbide,} {\protect\JournalTitle{Optica}} \textbf{6}, 991 (2019).

\bibitem{Vuckovic_4HSiC_nphoton}
D.~M. Lukin, C.~Dory, M.~A. Guidry, K.~Y. Yang, S.~D. Mishra, R.~Trivedi,
  M.~Radulaski, S.~Sun, D.~Vercruysse, G.~H. Ahn, and J.~Vu{\v c}kovi{\'c},
  \enquote{{{4H}}-silicon-carbide-on-insulator for integrated quantum and
  nonlinear photonics,} {\protect\JournalTitle{Nature Photonics}} \textbf{14},
  330--334 (2020).

\bibitem{Ou_4HSiC_combQ}
C.~Wang, Z.~Fang, A.~Yi, B.~Yang, Z.~Wang, L.~Zhou, C.~Shen, Y.~Zhu, Y.~Zhou,
  R.~Bao, Z.~Li, Y.~Chen, K.~Huang, J.~Zhang, Y.~Cheng, and X.~Ou,
  \enquote{High-{{Q}} microresonators on {{4H}}-silicon-carbide-on-insulator
  platform for nonlinear photonics,} {\protect\JournalTitle{Light: Science \&
  Applications}} \textbf{10}, 139 (2021).

\bibitem{Li_4HSiC_comb}
L.~Cai, J.~Li, R.~Wang, and Q.~Li, \enquote{Octave-spanning microcomb
  generation in {4H}-silicon-carbide-on-insulator photonics platform,}
  {\protect\JournalTitle{Photonics Research}} \textbf{10}, 870--876 (2022).

\bibitem{Ou_4HSiC_thermal}
X.~Shi, W.~Fan, A.~K. Hansen, M.~Chi, A.~Yi, X.~Ou, K.~Rottwitt, and H.~Ou,
  \enquote{Thermal {Behaviors} and {Optical} {Parametric} {Oscillation} in
  {4H}-{Silicon} {Carbide} {Integrated} {Platforms},}
  {\protect\JournalTitle{Advanced Photonics Research}} \textbf{2}, 2100068
  (2021).

\end{thebibliography}
\end{document}